# Is spacetime curved?
## Assessing the underdetermination of general relativity and teleparallel gravity

Ruward Mulder* & James Read†

**Abstract**

Realism about general relativity (GR) seems to imply realism about spacetime curvature. The existence of the teleparallel equivalent of general relativity (TEGR) calls this into question, for (a) TEGR is set in a torsionful but flat spacetime, and (b) TEGR is empirically equivalent to GR. Knox (2011) claims that there is no genuine underdetermination between GR and TEGR; we call this verdict into question by isolating and addressing her individual arguments. In addition, we anticipate and evaluate two further worries for realism about the torsionful spacetimes of TEGR, which we call the 'problem of operationalisability' and the 'problem of visualisability'.

# Contents

*University of Cambridge, Trinity College, CB2 1TQ, Cambridge, UK. ram202@cam.ac.uk.
†University of Oxford, Pembroke College, OX1 1DW, Oxford, UK. james.read@philosophy.ox.ac.uk.

# 1 Introduction

Realism about general relativity (GR) seems to imply realism about spacetime curvature. The existence of a version of teleparallel gravity—called the teleparallel equivalent of general relativity (TEGR)—calls this into question, since (a) TEGR is set in a torsionful but flat spacetime, and (b) TEGR is empirically equivalent to GR, so that no experiment could discern them.[1] So, building upon prior work by Lyre & Eynck (2003) and by Knox (2011), underdetermination and realism about GR warrant renewed attention.

The original version of teleparallel gravity was devised by Cartan (1922; 1923), and developed further by Einstein himself (1928b; 1928a; 1930) inspired also by his correspondence with Cartan (see their correspondence in (Debever 1979)). The key difference with GR is that it postulates an anti-symmetric connection instead of the (symmetric) Levi-Civita connection; the former is torsionful but flat, whereas the latter is torsion-free but curved. Einstein's aim was the unification of gravity with electromagnetism, as the latter could perhaps be captured by the six additional variables in the teleparallel theory—a project abandoned as physicists convinced themselves that the theory would not lead to novel predictions (Sauer 2006). Another hope of the unified field theory approach was to reproduce quantum effects, which likewise failed (Dongen 2010). A final large problem was the overdetermination of the field equations, resulting in practical obstacles for calculations. All this being said, teleparallelism has since been revived as an empirically equivalent alternative to GR; in this article, we will focus upon this modern-day version of the approach.[2]

Over the past decade, TEGR has attracted increasing attention from philosophers. Lyre & Eynck (2003) and Knox (2011) introduced into the philosophical literature the topic of the putative underdetermination between GR and TEGR; other recent works have discussed the 'background independence' of TEGR (Read 2023), the extent to which the theory really is empirically equivalent to GR once one considers boundary phenomena (Wolf and Read 2023), the status of equivalence principles in TEGR (Read and Teh 2022), and the existence of non-relativistic teleparallel theories (Read and Teh 2018; March, Wolf, and Read 2024; Wolf, Read, and Vigneron 2024). The underdetermination between GR and TEGR has also been discussed in (Wolf, Sanchioni, and Read 2024) (with regard to classic and modern tests of general relativity) and in (Fankhauser and Read 2023) (with particular focus on the results of gravitational redshift experiments). In the present article, we continue to explore this latter theme of the putative underdetermination between GR and TEGR, but now with a focus upon claims made by Knox (2011) to the effect that there is *no* genuine underdetermination between these theories.

The central aim of Knox' article is to undermine the underdetermination claim through a series of arguments that concern the difficulty of 'reading off' ontology from a theory or formalism. Unfortunately, Knox' article is not too explicit as to how we should identify such an ontology, except for inviting us to adopt a "relatively liberal attitude" (Knox 2011, p. 274). Here, we render Knox' arguments more explicit, but maintain—after careful analysis—that they are not conclusive: to follow Knox to her conclusion requires one to buy into certain philosophical and metaphysical commitments which might not be palatable to all—or which, at the very least, deserve to be rendered explicit so that one is in a better position to assess their palatability.

---

[1] Modulo, at least, issues to do with boundaries, discussed by Wolf & Read (2023), to which we return in Section 3.2.

[2] See (Aldrovandi and Pereira 2012) and (Bahamonde et al. 2023) for modern, book-length presentations of the theory.

In the second half of this article, we consider two further arguments—different from those raised by Knox—which might lead one to reject TEGR as offering a serious ontological alternative to GR. Both of these arguments have it that the concept of torsion is somehow metaphysically problematic. The 'problem of operationalisability' takes seriously the worry that torsion might not be as easy to operationalise as curvature. The 'problem of visualisability' articulates the question of whether torsion is harder to visualise than curvature. By extension, verdicts on that question might warrant thinking differently about the physical interpretation of TEGR as compared with that of GR. This runs into the question of whether intelligibility, perspicuity, visualisability, or the like, is necessary for the identification of the ontological commitments of a particular theory: questions which have recently been taken up in the philosophy of symmetries by e.g. Møller-Nielsen (2017), but which go back much further (Galison 1977; de Regt 1997; de Regt and Dieks 2005; Chang 2009). In response to this worry: we are optimistic about the prospects for visualising torsion. Not only is the non-closure of parallelograms (which constitutes the definition of torsion—more on this below) visualisable; in addition, there are visualisations of torsion available in the literature, such as certain three-dimensional crystalline structures in condensed matter physics.

Bringing all this together, we conclude that neither of these problems is sufficient to defuse the underdetermination; and since also Knox' arguments are at the very least questionable (in the literal sense), the challenge of underdetermination remains (at least in the absence of any other arguments which might defuse it)—thereby threatening the interpretation of curvature as a real property of spacetime.

However, that is not to say that GR/TEGR underdetermination cannot be defused. Although the current article focuses on how the above-mentioned defusing efforts are insufficient, other such efforts have in fact been proposed. For example, Wolf et al. (2024) argue that GR serves as a common core of the two theories (and of the 'geometric trinity of gravity' more generally—on which see (Beltrán Jiménez, Heisenberg, and Koivisto 2019)), since the Levi-Civita connection is always definable in each theory; separately, Mulder (2024a) argues that, by focusing on their common mathematical origin (the non-vanishing of the Lie bracket of covariant derivatives), the meaning of the terms 'torsion' and 'curvature' should be construed such that the distinction between them collapses.

The structure of this article is as follows. In Section 2, we recall the relevant background on TEGR. In Section 3, we present and evaluate Knox' arguments that there is no genuine underdetermination here. In Section 4 and Section 5, we address the problems of operationalisability and of visualisability of torsion, respectively.

# 2 Formalism of teleparallel gravity

'Teleparallel gravity' is an umbrella term for a set of spacetime theories on parallelisable manifolds which deploy an anti-symmetric compatible affine connection, as opposed to the Levi-Civita connection of GR.[3] Instead of exhibiting spacetime curvature, teleparallel theories manifest spacetime *torsion*, which is nothing other than a codification of the antisymmetry of an affine connection $\Gamma$, as encoded in the 'torsion tensor',[4]

$$T^{\tau}{}_{\mu\nu} := \Gamma^{\tau}{}_{\mu\nu} - \Gamma^{\tau}{}_{\nu\mu}. \tag{2.1}$$

Geometrically, torsion captures the extent to which two infinitesimal vectors fail to commute when parallel transported along themselves—pictorially, this is often glossed as the 'non-closure of infinitesimal parallelograms'.

Turning to dynamics, Hayashi & Shirafuji (1979) give one unified Lagrangian for all teleparallel theories, depending upon which value one takes for three independent parameters, $a_1$, $a_2$, $a_3$ (and a constant $\kappa$):

$$\mathcal{L} = \kappa \left( a_1 T_{\mu\nu}{}^{\rho} T^{\mu\nu}{}_{\rho} + a_2 T_{\rho\nu}{}^{\mu} T_{\mu\nu}{}^{\rho} + a_3 T^{\rho}{}_{\mu\rho} T_{\nu}{}^{\nu\mu} \right). \tag{2.2}$$

This is what was meant above with 'teleparallel gravity' being an umbrella term. The teleparallel equivalent to general relativity (TEGR) is obtained by taking $a_1 = 1/4$, $a_2 = 1/2$, $a_3 = -1$,

$$\mathcal{L}_{\text{TEGR}} = \kappa \left( \frac{1}{4} T_{\mu\nu}{}^{\rho} T^{\mu\nu}{}_{\rho} + \frac{1}{2} T_{\rho\nu}{}^{\mu} T_{\mu\nu}{}^{\rho} - T^{\rho}{}_{\mu\rho} T_{\nu}{}^{\nu\mu} \right), \tag{2.3}$$

which can then be rewritten as

$$\mathcal{L}_{\text{TEGR}} = \kappa \mathcal{T}, \tag{2.4}$$

where

$$\mathcal{T} := \frac{1}{4} T_{\mu\nu}{}^{\rho} T^{\mu\nu}{}_{\rho} + \frac{1}{2} T_{\rho\nu}{}^{\mu} T_{\mu\nu}{}^{\rho} - T^{\rho}{}_{\mu\rho} T_{\nu}{}^{\nu\mu} \tag{2.5}$$

is the 'torsion scalar'. This Lagrangian can in turn be rewritten in terms of the Ricci scalar $\mathcal{R}$ and a boundary term:

$$\mathcal{T} = -\mathcal{R} - 2\nabla^{\nu} T_{\rho\nu}{}^{\rho}. \tag{2.6}$$

This boundary term does not contribute to the dynamics and thus establishes—at least at the dynamical level—the equivalence with the Einstein-Hilbert action and hence the Einstein equations.[5]

There is only one further technical point to make which will be necessary for our purposes in this article. Can one say more about the particular form that the torsionful (but flat) derivative operator will take in TEGR, analogous to the form of the Levi-Civita connection takes in terms of the metric tensor in GR? Denoting the coefficients of a completely anti-symmetric connection $\nabla$ as $\Gamma^\rho{}_{\mu\nu}$ and the components of the (completely symmetric) Levi-Civita connection $\mathring{\nabla}$ for a metric $g_{\mu\nu}$ as $\mathring{\Gamma}^\rho{}_{\mu\nu}$, one can show that these are related as[6]

$$\mathring{\Gamma}^\rho{}_{\mu\nu} = \Gamma^\rho{}_{\mu\nu} - K^\rho{}_{\mu\nu}, \qquad (2.7)$$

where $K^\rho{}_{\mu\nu}$ is the 'contorsion tensor', which is defined in terms of the torsion tensor as

$$K_{\mu\nu\lambda} = \frac{1}{2}\left(T_{\mu\nu\lambda} + T_{\nu\lambda\mu} - T_{\lambda\mu\nu}\right). \qquad (2.8)$$

---

[3] Recall that a differentiable manifold $M$ is parallelisable if and only if it admits a set of smooth vector fields $\{V_1, \ldots, V_n\}$ such that, at every point $p \in M$, the tangent vectors $\{V_1(p), \ldots, V_n(p)\}$ form a basis of the tangent space $T_p M$.

[4] We will generally use coordinate indices in this article rather than abstract indices.

[5] See (Wolf and Read 2023) for subtleties regarding this equivalence claim when boundary phenomena are taken into account, where it is argued that even though the two theories are dynamically equivalent, having actions which differ by a boundary term does have empirical consequences.

Sometimes the TEGR connection is referred to as 'Weitzenböck connection'—we will follow suit in this article, and moreover will in general work within the 'pure tetrad' framework according to which the components of this connection can be written exclusively in terms of the tetrad fields $e^\mu{}_a$ (also called 'tetrads', 'frame fields' or 'vielbeins') and their inverse 'co-tetrads' $e^a{}_\mu$, as[7]

$$\Gamma^\rho{}_{\mu\nu} = e^\rho{}_a \partial_\nu e^a{}_\mu. \qquad (2.9)$$

There is, however, a 'covariant' approach to TEGR, in which the spacetime connection has components which are invariant under local Lorentz transformations—see (Aldrovandi and Pereira 2012, pp. 11–2, 48–9). We return to this distinction in Section 3.2; in the meantime, see (Aldrovandi and Pereira 2012; Bahamonde et al. 2023) for further background on the formalism of TEGR.

## 3 Knox on teleparallel gravity

With the relevant formalism of TEGR in hand, we turn now to an assessment of Knox' (2011) claims to the effect that there is no genuine underdetermination between GR and TEGR, because the ontology of TEGR is 'really' that of GR. The word 'force' in the title of Knox' article—'Newton-Cartan theory and teleparallel gravity: the force of a formulation'—should then be read with a pejorative ring to it: it signals that TEGR as a mathematical formulation has an ideological force behind it that fools us into thinking that we are confronted with a genuine example of underdetermination of theory by data. Knox does not argue against such underdetermination by *disputing* the empirical equivalence of these two theories; on the contrary, she argues *for* their ontological equivalence, in the sense that they should be interpreted as making the same claims about what nature is like[8]—that the theories are empirically equivalent thereby (for Knox) follows straightforwardly. On this, Knox writes:

> [...] these theories do not, in fact, represent cases of worrying underdetermination. On close examination, the alternative formulations are best interpreted as postulating the same spacetime ontology. In accepting this, we see that the ontological commitments of these theories cannot be directly deduced from their mathematical form. (Knox 2011, p. 264)

---

[6] One way to do this is to repeat the standard derivation of the Levi-Civita connection coefficients but allowing for both symmetry and anti-symmetry, as follows. Write the metric compatibility condition three times, cycling through the indices, and add these together as: $0 = \nabla_\rho g_{\mu\nu} - \nabla_\mu g_{\nu\rho} - \nabla_\nu g_{\rho\mu}$. Writing out the connection coefficients, and using the definition of the torsion tensor (2.1) and its anti-symmetry property $T_{\mu(\nu\rho)} = 0$, will result in (2.7).

[7] The tetrads cover the manifold such that for each point there is a set of four independent vector fields that span the tangent bundle at that point: $e_a = e_a{}^\mu (\partial/\partial x^\mu)$, for $a = 0, 1, 2, 3$. Co-tetrads are then $e^a = e^a{}_\mu dx^\mu$, so that $e_a{}^\mu e^b{}_\mu = \delta_a{}^b$, and these connect to spacetime vectors as $e_a{}^\mu V^a = V^\mu$.

At the same time, in Knox' eyes these two equivalent formulations are not on a par: representations from GR take precedence, for her, over those from TEGR. It is not that GR is a reformulation of TEGR—rather, loosely, the situation is more akin to TEGR being a bad 'coordinate choice' for GR. That is, in the sense that it is an unilluminating way of splitting up degrees of freedom (cf. the relationship between Newtonian gravity and Newton-Cartan theory, as also discussed by Knox (2014)).

The points in Knox' article are important for bringing together philosophical themes concerning TEGR and for presenting the theory in a manner which is accessible to philosophers. That work was long overdue. Moreover, Knox furthers the debate by clearing up (at least in part) some important muddles, such as TEGR being regarded as a 'force' theory akin to Yang-Mills theory (see also (Wallace 2015)) and, importantly, drawing attention to how non-gravitational dynamics can influence our interpretation of the gravitational sector.[9] That being said, Knox' main argument—that TEGR should be seen as a reformulation of GR instead of as a rival—does not move us, as it rests on questionable assumptions which remain unarticulated. And as we will show, these assumptions appear to stem from prior familiarity with GR; to stem from an Occamist drive to expunge gauge degrees of freedom at all costs; from a commitment to an understanding of what is meant by 'spacetime structure' which is problematically theory-independent; and from certain other implicit philosophical commitments, such as a specific form of spacetime functionalism (more on which later). Below, we present and assess Knox' critiques of TEGR in the form of three 'problems' that are to be taken as potential motivations to prefer GR. We argue that in each case, Knox' objections to TEGR are relevant but not decisive.

---

[8]We believe other qualifications, like 'theoretical equivalence' or 'formal equivalence' would be less apt than 'ontological equivalence' due to their mathematical connotations. Weatherall & Meskhidze (2024), for example, argue that GR and TEGR are not theoretically equivalent (in the specific sense that they are not categorically equivalent).

[9]Regarding the latter, we do commend Knox' discussion that considering *non-gravitational* forces and their associated force-free trajectories can make a difference in evaluating the seriousness of the underdetermination. She applies this to the non-relativistic case of putative underdetermination between Newtonian gravity and Newton-Cartan theory. Yet, the opposite can also be the case: in Weyl's 'geometrisation' of the electromagnetic field (e.g., Giovanelli 2021), for example, particles follow the affine geodesics of different affine connections, namely those corresponding to different values of electric charge. Furthermore, if such characteristics are deemed problematic, sometimes other formulations can circumvent them: in five-dimensional Kaluza-Klein theory, different values of charge are identified with different values of momentum in an additional spatial dimension. The point is that when one adds non-gravitational forces to break underdetermination, the choice about how to add these forces mathematically is itself subject to interpretation.

## 3.1 Reverse-engineering the metric and connection of GR

One of Knox' main accusations against TEGR is that all the features of the theory are such that they in fact reproduce the familiar geometrical properties or objects of GR. Thus, the implication is that TEGR is somehow parasitic on GR and its ontology. Knox' main examples are (i) the obscured use of the metric, and (ii) the coupling of matter to the Levi-Civita connection. We look at these in turn.

Knox argues that the metric tensor of GR is obscured in TEGR and yet remains a central tool in that theory:

> So far we have not mentioned the metric. In teleparallel gravity, this is quite possible; it does not appear in the formalism of the theory. Nonetheless, it is worth noticing that it has been hiding in the shadows all along, closely tied to the tetrad field. In fact, $g_{\mu\nu}$ is still used to raise and lower indices, just as it is in GR.
>
> One might therefore have doubts that teleparallel gravity really postulates a different ontology; the old entities from GR appear to be waiting in the wings. (Knox 2011, pp. 273–4)

As mentioned in the previous section, it is indeed quite typical to see TEGR presented in terms of tetrads, for we have already seen that these fields can be used—via (2.9)—to construct the Weitzenböck connection coefficients in TEGR; moreover, (co-)tetrads are related to the metric as

$$g_{\mu\nu} = e_\mu{}^a e_\nu{}^b \eta_{ab}, \tag{3.1}$$

for all spacetime points $p \in M$ and the tangent space Minkowksi metric field $\eta_{ab}$. With this in mind, we can then formulate a potential problem for TEGR:

> **Problem 1(a): The shy metric tensor:** TEGR is usually set in the tetrad formalism and its main quantities are commonly expressed in terms of those fields. Yet, the metric is still 'waiting in the wings', so to speak, not only through (3.1) but also because it is used to raise and lower spacetime indices. Since TEGR relies on the metric tensor in this way, we should therefore not read off the ontology of the theory from its surface formalism.

How should we respond to this problem? It is indeed true that the GR metric as given in (3.1) is used to raise and lower spacetime indices in TEGR. Of course, one might argue that were one to use only the right-hand side of this equation to raise and lower indices, then reference to the metric tensor would thereby be avoided. Such a reply, however, might be too quick: it is the same piece of structure that is doing the raising and lowering and this piece of structure is a complex of two tetrads whereas in GR it is the metric tensor directly, part and parcel of its conceptual framework.

But, in any case, there is no obligation to set up TEGR using tetrad fields: tetrads are generally used for convenience in order to express relevant physical quantities in the theory, but this is by no means compulsory.[10] Indeed, there are metric formulations of TEGR just as well as there tetrad formulations. Up to a boundary term, the TEGR Lagrangian (2.3) can be coupled into an action as

$$S^g_{\text{TEGR}} = \int dx^4 \sqrt{-g}\mathcal{T} \tag{3.2}$$

just as well as it can be written as

$$S^e_{\text{TEGR}} = dx^4 \int e\mathcal{T}, \tag{3.3}$$

where $g$ is the determinant of the metric tensor, cf. (Hohmann 2021; Capozziello, Falco, and Ferrara 2022). Conversely, tetrads can also be introduced freely while using GR and indeed this is routinely done (Møller 1961a,b, 1978). Table 1 summarises these four possible sectors. Perhaps one could still argue as follows: for the metric formulation of TEGR, we are actually even closer to GR than in the tetradic formulation. However, even in this case one would be ontologically committing to not only the metric but also an anti-symmetric connection (and hence not the Levi-Civita connection). The case remains that we can formulate either TEGR or GR in terms of tetrads or in terms of the metric tensor (plus spacetime derivative operator). Thus one can only conclude that the use of tetrad fields is orthogonal to working within the framework of TEGR, and so critiques which invoke said formalism in an essential way cut no ice.

---
[10]Tetrads are also helpful for understanding TEGR as a 'gauge theory of the translations'—see (Aldrovandi and Pereira 2012).

That being said, even if we were to restrict ourselves to a version of TEGR which makes use of tetrads and a version of GR which does not, we think that Knox' argumentation could be questioned. Within that context, Knox still looks at TEGR and takes the metric to be more fundamental than the tetrad fields:[11]

> Both [TEGR] and GR appear to take the metric to be fundamental. Looked at another way, we can note that both theories admit of both the tetrad field and the metric; tetrad formulations of GR have various uses. Moreover, rods and clocks survey the self-same metric in both theories. The only difference is the way in which this comes about—either via the Weitzenböck connection or the Levi-Civita connection. It seems that both theories posit the 'same' spacetime; if the connections in the two theories are thought of as modelling properties of this spacetime, they should perhaps be seen as alternative representations of the same properties. (Knox 2011, p. 273)

But the fact that the metric and tetrad formulations of either theory are intertranslatable does not tell us anything about the relative 'fundamentality' of the metric or tetrads.[12] Nor does the observation that metrical structure is just as well oper-ationalised in both theories—by linking metrical structure to lengths and durations surveyed by rods and clocks—*per se* privilege the metric over the tetrads.[13]

---

[11]Here, neither we nor Knox (2011) should be understood as taking 'fundamental' to be synonymous with 'real'. Moreover, while of course all parties could say more about what 'fundamentality' in physics really amounts to (see e.g. (Tahko 2023)), these issues don't really matter for the purposes of the discussion here; one can simply choose to work with one's preferred account of the notion.

[12] In general, there is no clear-cut reason why either the tetrad formulation or the metric formulation regarded as being more apt to represent reality. Usually the metric degrees of freedom are taken to be sufficient to capture the real degrees of freedom. It is of course important (fn. 3) that the manifold

Indeed, the difference between GR and TEGR just is the use of a symmetric versus an anti-symmetric connection, with the corresponding consequences for curvature and torsion. It is the realistic interpretation of curvature and torsion as properties of spacetime that is in the balance, while we already know that there is no discrepancy of measurement expected for metrical properties. So: if Knox indeed has a reason for privileging the formalism of GR over that of TEGR, then this likely has to do with some other considerations, such as those discussed in Section 3.2 and Section 3.3.

Ultimately, it is not so surprising that metrical structure can be constructed in TEGR. After all, in TEGR (just as in GR) we need to have measures of length and duration, as surveyed by rods and clocks. This metrical structure can be encoded either by the metric tensor or by the tetrads. But the point of metric-affine geometries such as TEGR is that the metric structure and the affine structure are *not* as intimately tied up with each other as they are in GR. In GR it is common to regard the metric as ontologically prior to the connection, since the former uniquely determines the latter. But this interpretation is not necessary and it may even be more helpful to regard the connection as ontologically prior—a point made, for example, by Stachel (2007).

In TEGR, the conceptual cleavage of metrical structure (whether represented by a metric tensor or by tetrads) and affine structure (represented by the anti-symmetric connection) is more readily recognised. The salient difference between GR and TEGR has to do with the symmetry of the connection, and (in the quote above) Knox (2011, p. 273) points exactly at this: "[...] rods and clocks survey the self-same metric in both theories. The only difference is the way in which this comes about—either via the Weitzenböck connection or the Levi-Civita connection." The question is not about whether the metric is ontologically prior to either the tetrads or the connection, but rather about whether the different connections are to be *interpreted* as giving rise to genuinely different spacetime structures (*in casu*: curved or torsionful spacetimes, and thus whether curvature or torsion, if any, is deserving of a realistic interpretation as a property of spacetime).

We take the above to dispatch worries regarding the metric in TEGR, at least conditional on clearing up some issues regarding scientific realism to which we will later return. This leaves us with (ii), concerning the putative reverse-engineering of the Levi-Civita connection. This is more involved, because it is tied closely to a worry

---

be parallelisable, i.e. admit of tetrads in the first place. To the extent that this is an objection, it applies to tetrad formulation of GR as well as to tetrad formulations of TEGR (as also pointed out by Knox (2011), fn. 29, p. 272). Nevertheless, such restrictions can be seen as very natural from either standpoint, and in fact could be argued to constitute a type of evidence for tetrads (cf. (Nawarajan and Visser 2016)); we thank an anonymous referee for this point. Another reason for taking tetrads seriously is that they—unlike the metric—admit of fermionic degrees of freedom by allowing for the construction of Dirac operators in terms of the tetrads, cf. (Krasnov 2020).

[13] Although of course, due to their extra gauge freedom, there is a clear sense in which the tetrads are less operationalisable than the metric field. The situation here is somewhat akin to debates regarding the metaphysics of electromagnetism—in principle, both *F*-field and *A*-field electromagnetism offer viable metaphysics for the theory, in spite of the latter having more gauge freedom and thereby being less operationalisable.

|                          | metric formalism              | tetrad formalism       |
|--------------------------|-------------------------------|------------------------|
| **symmetric connection** | $\int dx^\mu \sqrt{-g}\mathcal{R}$ | $\int dx^\mu e\mathcal{R}$ |
| **anti-symmetric connection** | $\int dx^\mu \sqrt{-g}\mathcal{T}$ | $\int dx^\mu e\mathcal{T}$ |

Table 1: Actions for both general relativity and teleparallel gravity can be formulated either in terms of metrics or in terms of tetrads.

about inertial structure which we address in Section 3.3, but we nevertheless discuss it here as a conceptually independent point of reverse-engineering.

To establish the existence of conserved quantities in TEGR, one uses the Levi-Civita connection for the minimal coupling rule, re-expressing the Weitzenböck connection $\nabla$ of TEGR in terms of the Levi-Civita connection of GR and the contorsion tensor. That is, the field equations of TEGR lead to a conserved quantity $hj_a{}^\mu := (\partial \mathcal{L}_{\text{TEGR}}/\partial e^a_\mu)$ only with the *teleparallel covariant connection*, which—one may argue, looking at (2.7)—is the Levi-Civita connection in disguise (in the sense that, ultimately, this is a GR expression simply rewritten in TEGR quantities):

$$D_\mu j_a{}^\mu := \partial_\mu j_a{}^\mu + \left(\Gamma^\mu{}_{\nu\mu} - K^\mu{}_{\nu\mu}\right) j_a^\nu. \tag{3.4}$$

This mimicking can be formulated as another problem:

> **Problem 1(b): Reverse-engineering the Levi-Civita connection:** In TEGR the teleparallel covariant derivative (3.4) is made up out of the contorsion tensor and the Weitzenböck connection, thereby closely mimicking the Levi-Civita connection of GR. It would be more natural were the conserved quantities to be derived using the Weitzenböck connection. Hence, it seems that it is the Levi-Civita connection of GR which is doing much of the physical work in TEGR.

This problem seems to have real bite, since it highlights that—at least when it comes to conserved quantities—the Weitzenböck connection might not play as central a role in TEGR as does the Levi-Civita connection.

In response to the worry, we say the following: versions of teleparallel gravity are plentiful, not only through different choices for the parameters of the Hayashi-Shirafuji Lagrangian (2.2) but also through the *choices* that are made to couple fields dynamically. It is true that TEGR is constructed explicitly so as to reproduce the empirical content of GR; indeed reverse-engineering is to some extent necessary to preserve the empirical equivalence with GR. Therefore it is only to be expected that teleparallel quantities can always be rewritten in a form that uses only quantities associated with GR.

Does this constitute a genuine problem? We think not. In principle TEGR can stand on its own four legs—the geometric structure (whether cashed out in terms of tetrads or in terms of a metric field) and the Weitzenböck connection—even if GR had never been invoked.[14]

---

[14]In response to this, Knox (2011) could perhaps appeal to additional arguments against geometric

To emphasise: reverse-engineering by itself does not strengthen the claim that TEGR provides us with just a different mathematical form to represent the same observables that GR represents. It is simply a statement of symmetry. As such, it goes both ways: any quantity in GR could be seen as reverse-engineered from TEGR, with all the teleparallel quantities "hiding in the shadows all along". For, one may equally well have a *general relativistic covariant connection*, analogous to the teleparallel covariant connection (3.4),

$$\mathring{D}_\mu j_a{}^\mu := \partial_\mu j_a{}^\mu + \left(\mathring{\Gamma}^\mu{}_{\nu\mu} + K^\mu{}_{\nu\mu}\right) j_a^\nu, \tag{3.5}$$

which says that it is the Levi-Civita connection together with the contorsion tensor that are closely mimicking the Weitzenböck connection. Again, this is simply a statement of symmetry.

Albeit not for the reason of reverse-engineering, this symmetry between the theories might nevertheless not make them entirely on a par: the symmetry-related quantities do not stand in a one-to-one but in a many-to-one relation. That is, for each model of GR, there will generically be many models of TEGR (at least when presented in the tetrad formalism). The additional degrees of freedom in TEGR are considered unphysical or gauge, which brings us to the problem to be addressed in Section 3.2.

We also highlight another problem down the line: the coupling to matter using the teleparallel covariant connection can be construed as a problem insofar as one sees it as more than a mathematical formulation and rather as identifying directly the inertial structure which the theory must represent; we return to this in Section 3.3.

## 3.2 Surplus structure

Setting aside issues to do with the fact that the quantities of GR can be expressed in terms of those of GR and *vice versa*, we turn now to a second conceptual problem with TEGR adumbrated by Knox—namely, that this theory has surplus ('gauge') structure as compared with GR. Here is Knox:

> The tetrad field is defined only up to a local gauge transformation, and this gauge freedom passes on to the connection defined in terms of the tetrads. As a result, the gravity/inertia split expressed in [(2.7)] is just as much a gauge matter as it was in the Newtonian case. In fact, the situation is worse. Because the tetrad field is subject to a local gauge freedom, no amount of information about the tetrad field on, say, a spacelike hypersurface will determine the value of the tetrad field in other regions of

---

reformulation of this kind presented by Dürr & Ben-Menahem:

> Else, it's too cheap to concoct geometric alternatives [...], e.g. by introducing empirically and theoretically superfluous extra structure. [Footnote suppressed.] Instead [...], the differences of geometric alternatives must be non-trivial (e.g. simplifying, explanatory, unificatory, etc.) (Dürr 2022, p. 171)

Our thanks to an anonymous reviewer for drawing our attention to this passage; however, they go beyond the arguments offered in (Knox 2011).

spacetime. As a result, it is tempting to think that the metric, and the Levi-Civita connection, should be taken as ontologically prior to the tetrad field and Weitzenböck connection. (Knox 2011, p. 273)

There are two related problems here. First, any model of tetradic TEGR in which the tetrad fields are related by a local Lorentz transformation $e_\mu{}^a \mapsto \Lambda^a{}_b(x) e_\mu{}^b$ will also be a TEGR model associated to the same GR model. Indeed, whereas the metric tensor (which is symmetric) has $\frac{4(4+1)}{2} = 10$ independent components, two tetrad fields have an additional six: $4 \times 4 = 16$. Second, even though it is true that "this gauge freedom passes on to the connection defined in terms of the tetrads", this is only true on the conception of TEGR on which the theory deploys a Weitzenböck connection; *not* on the covariant conception of TEGR which uses a 'dressed' connection. So, setting aside the second of these two issues for the time being, here is the problem:

> **Problem 2: Surplus degrees of freedom**: There are surplus ('gauge') degrees of freedom in TEGR that are not present in GR, recognised in (3.1). Thus, for each model of GR there are many empirically equivalent models of TEGR.

Potentially, this complaint is appropriate, at least on grounds of parsimony: let us not take these mathematical variables seriously (one might say) when they do not lead to additional empirical predictions, for adding something on top of an already empirically adequate theory is excessive (notwithstanding any calculational or other pragmatic advantages).

Other grounds would involve concerns with indeterminism, as indeed Knox is pointing at in the above quote when she says that the specification of initial values on spacelike hypersurfaces fails to fix the tetrad at other times. Philosophically, this is akin to the hole argument in GR (Earman and Norton 1987; Pooley 2013; Gomes and Butterfield 2023): if substantivalism is understood as the position that spacetime points have an intrinsic 'thisness' (or haecceity), then due to diffeomorphism invariance no amount of information on a spacelike hypersurface will determine the distribution of spacetime points in other regions of spacetime, amounting to indeterminism.

Take as another example the different formulations of electrodynamics, one in terms of the electric and magnetic fields $(\mathbf{E}, \mathbf{B})$ and one in terms of the electromagnetic potentials $(\phi, \mathbf{A})$. In the latter case, Maxwell's equations formulated in terms of the potentials will generically not have unique solutions (Belot 1998). This is not a problem as long as one interprets the additional mathematical variables (additional to those shared with the electric and magnetic field formulation) as unphysical. As such, it only leads to indeterminism if one has a reason to regard these additional variables as having physical significance of their own. Yet, for the gauge variables in TEGR, there is no such reason;[15] for the gauge variables of the potentials theory $(\phi, \mathbf{A})$, the reason is the empirical phenomenon known as the Aharonov-Bohm effect, and the resulting situation of having no experimental means to find out *how much* of the (hitherto) gauge degrees of freedom should be considered physical can be characterised as 'gauge-underdetermination' (Mulder 2021). Thus, indeterminism only arises when 'gauge' variables are considered to have physical significance after all—a move that is unnecessary in TEGR. This leaves us only with the argument that the additional gauge degrees of freedom in TEGR are simply superfluous.

---

[15]At least as far as we are aware, and nor does Knox mention such a reason.

Perhaps, however, the Occamist drive itself—to excise 'gauge' structure—should be tempered, or at least recognised to be suffixed with important *ceteris paribus* clauses. One reason would be that were one to start out with the formalism of TEGR, and then to expunge the gauge variables, this does not immediately lead to a formalism ready to be identified with GR. After all, there remains the non-trivial translation (2.7) between the flat but torsionful connection and the Levi-Civita connection. This is strikingly different from the situation in electrodynamics, where expunging the gauge freedoms of the potentials leaves one with a structure that can be directly identified with the electric and magnetic field.[16]

Another reason to resist the philosopher's drive to excise gauge variables might very well be that it is in some cases naïve to think it will not have empirical consequences: 'gauge' degrees of freedom are now well-understood to be required for, among other things, coupling subsystems (Rovelli 2014; Teh 2015; Gomes 2021), and the correct modelling of boundary degrees of freedom (Murgueitio Ramírez and Teh 2022; Wolf, Read, and Teh 2023). Indeed, in (Wolf and Read 2023) the merits of TEGR over GR with respect to modelling boundary phenomena are highlighted explicitly. Therefore, it is far from clear on physical grounds that the extra degrees of freedom in TEGR render the theory conceptually problematic to the degree that there is no genuine underdetermination between it and GR, as Knox would have it.

## 3.3 Inertial structure and functionalism

Standardly, necessary conditions for a frame to be *inertial* are that it be one in which the connection coefficients vanish and on which the coordinate axes remain unchanged by parallel transport, i.e., when transported over the straightest lines of that connection. According these criteria, inertial structure would seem to be a notion relative to the choice of connection. As Knox (2011, p. 268) sees it, such a definition importantly already assumes that we have picked the *correct* connection, the one that conveys the physical inertial structure. Knox continues:

> Happily, there is a more physical characterisation of an inertial frame available. Inertial frames are those reference frames in which force-free bodies move with constant velocities, and which are indistinguishable according to the dynamics. This second criterion is a strong one; in order for a class of reference frames to count as inertial, dynamical laws must take the same form in each frame. Moreover, inertial frames must be universal; if the

---

[16] See (Belot 1998) for a more detailed discussion of the gauge orbits of electrodynamics; and see again (Mulder 2021) for a more detailed discussion of how such expunging leads to a *widening of equivalence classes* and how such expunging in the form of gauge-fixing is subject to choices of interpretation, even though empirically constrained by the Aharonov-Bohm effect.

theory under consideration allows for multiple types of interaction, each of these must pick out the same class of inertial frames. (Knox 2011, p. 268)

Implicit in this article is the by-now well-known philosophical position of 'spacetime functionalism', which has it that "the spacetime role is played by whatever defines a structure of local inertial frames" (Knox 2018, p. 122).[17] In turn, as discussed in (Knox 2011, p. 268) and (Knox 2013, p. 348), the structure of local inertial frames is picked out by any structure which identifies the frames of reference in which (a) force-free bodies move with constant velocities, (b) the dynamics governing material bodies simplifies maximally, i.e., the laws take the same form and (c) such is the case for all material bodies and associated dynamics ('universality').

Given this version of spacetime functionalism, it would seem to follow that the spacetime structure of TEGR *just is* that of GR, for (i) being torsionful, the components of the Weitzenböck connection cannot be made to vanish at a point, and (ii) it would seem that it is the Lorentzian metric of GR and its associated Levi-Civita connection which play the spacetime role in TEGR (Knox 2013, p. 347).[18] We condense this problem for TEGR as follows:

> **Problem 3: Non-vanishing connection coefficients:** The structure of inertial frames is specified by requirements (a)–(c). Being torsionful, the components of the Weitzenböck connection cannot be made to vanish in any frame at any point $p \in M$. Therefore, the Weitzenböck connection cannot pick out a structure of local inertial frames, and so cannot (by spacetime functionalism) be the correct spacetime setting for TEGR.

As outlined above, Knox indeed goes on to argue that the true spacetime of TEGR is that of GR—i.e., the familiar Lorentzian spacetimes of that theory—because material bodies couple to that metric and its associated Levi-Civita derivative operator.[19]

There are various options to respond to this apparent problem for the viability of a literalist interpretation of TEGR *qua* spacetime theory; here we focus on the following three:

1. Reject Knox' spacetime functionalism.

2. Reject Knox' functional characterisation of inertial structure.

3. Deny the 'realiser functionalism' which is implicit in Knox' reasoning.

Let us discuss each of these in turn, beginning with option (1). Frankly, insofar as one is motivated to identify (or individuate) a structure which plays the 'spacetime role' in a given physical theory, we in fact find Knox' functionalism attractive.[20] Nevertheless, surely one won't go to jail for not buying into Knox' particular brand of spacetime functionalism, and indeed other options are available—see (Lam and Wüthrich 2018; Baker 2020). Focusing here on the approach of Baker (2020), we see that other authors have proposed that 'fundamentality' should constitute part of the spacetime role. Yet, to us there seems to be nothing wrong with insisting that it is (say) the tetrads or Weitzenböck connection which are fundamental in TEGR—in which case we see that alternative functionalist criteria might judge that such objects are the spatiotemporal structures of the theory after all. With this we do not mean to *endorse* Baker's reasoning—we simply invoke it as a means of pointing out that it is certainly possible to push back against Knox' functionalist argument.

---

[17] In personal communication, Knox has indicated to us that she understands (Knox 2011) not as *presupposing* spacetime functionalism, but rather as an *argument for* it. But since spacetime functionalism does appear to be operative in the above passage, we have some worries about circularity in that case.

[18] One can in fact find 'anholonomic' frames in which the components of torsionful connections vanish; for discussion and critical assessment of the physicality of such frames, see (Knox 2013, p. 354). We won't engage further with anholonomic frames in this article.

[19] Knox' account is reconstructed in, e.g., (Baker 2020; Read and Menon 2021).

Moving on to possible response (2), Knox' criterion that inertial structure should be associated with inertial *frames*, and so with vanishing connection coefficients, sounds reasonable at first blush. However, one might push back against this for both technical and conceptual reasons. Technically: making good mathematical and physical sense of 'local inertial frames' in a given spacetime theory is by now very well-known to be a delicate issue, as testified by (Read, Brown, and Lehmkuhl 2018; Fletcher 2020; Weatherall 2020; Fletcher and Weatherall 2023a,b; Linnemann, Read, and Teh 2024), among others. Again we agree with Knox that the notion of a local inertial frame, or of the local validity of special relativity in GR, is meaningful and comprehensible, but we will refrain from dwelling on this point in the current paper; our point here is simply that there is room for other positions, and indeed that several authors may in fact plump for such alternative positions at exactly this juncture.

Conceptually: it is not obvious why the vanishing of connection coefficients should be regarded as being the *sine qua non* of inertial structure—for it is of course still true of the Weitzenböck connection (as with any other affine connection) that it identifies a preferred class of trajectories as the affine geodesics, which are thereby naturally associated with straight-line motion. Why should one dismiss this obvious (and, indeed, well-defined and unambiguous!) notion of inertial structure in favour of an account in terms of the vanishing of connection coefficients?

Now, it is true that Knox stresses in other work that she does not intend to offer a once-and-for-all account of inertial structure:

> One might, if one wished, take these features to be something like 'constitutive' of our concept of an inertial frame. But I don't mean this to imply that there is something like an inertial frame concept that may be defined once and for all and is common to all spacetime theories. Concepts in physics vary in subtle ways from theory to theory and from application to application [...] Inertial frames in general relativity do differ from those in, say, special relativity, most notably because they are only defined locally. Nonetheless, it's also highly non-trivial that so many of the features of an inertial frame are retained in general relativity. (Knox 2013, fn. 4)

---

[20]Although one might want to seek a yet-more 'operationally informed' variant *à la* Read & Menon (2021).

Fair enough—but why then carry over Knox' conception of an inertial frame to other theories, such as TEGR? Or to theories yet further removed from GR? Clearly, more needs to be said to justify the domain of applicability of Knox' account.[21]

Finally, we discuss option (3). We remind the reader of the quote from the introduction, where Knox writes *vis-à-vis* GR and TEGR that

> [...] these theories do not, in fact, represent cases of worrying underdetermination. On close examination, the alternative formulations are best interpreted as postulating the same spacetime ontology. In accepting this, we see that the ontological commitments of these theories cannot be directly deduced from their mathematical form. (Knox 2011, p. 264)

In making such claims, Knox moves from the *individuation* of structures with the same functional role in models of the two theories—structures qualifying as spatiotemporal in light of her functionalist criterion—to a claim that those structures should be *identified* as the very same ontology. This position is sometimes known as 'realiser functionalism' (Cohen 2005; McLaughlin 2006; Levin 2018)—indeed, in the context of Knox's spacetime functionalism, this point has already been made by Lam and Wüthrich (2020, p. S344). This, of course, is a non-trivial position which not everyone need accept (cf. the status of realiser functionalism in the philosophy of mind).[22]

## 4 Operationalisability

In this section, we consider and assess a potential objection which could be raised against TEGR and other torsionful theories: that such theories are deficient because torsion is not an operationalisable geometrical notion. Before we get to this objection, though, we must clarify what it means for a piece of structure in a physical theory to be operationalisable.

Roughly, for a piece of structure (or concept) to be operationalisable is for it to be coordinated in a systematic way with an empirical/observable process. On this understanding, operationalisability is weaker than measurability, which requires in addition that one be able to find experimental setups which allow agents to gain epistemic access to that structure, given sufficiently warranted background assumptions.[23] On this understanding, the curvature tensor of GR can be said to be *operationalised*

---

[21]Here is a further point about specifying a functional role in terms of the theory itself, or in some theory-transcendent way, using a meta-language that is based on other theories or familiarity with other theories, or on the manifest image. To us, it seems epistemically more cautious to identify a concept (such as 'inertial frame') from the point of view of a theory itself. Any theory-transcendent criteria should be independently motivated—Knox, of course, can be seen to do this elsewhere (Knox 2013, 2014, 2018).

[22]In the spirit of keeping the semantic and epistemic conjuncts of scientific realism apart, note that, analogously to the distinction between *structuralism* and *structural realism* in many cases it is useful to distinguish between functionalism and can be called *functional realism*—see (Mulder 2024b, p. 11). Where structuralism and functionalism are purely semantic tools for describing or defining a certain piece of structure or set of items, the realisms go beyond this semantic individuation and come with a commitment to the *existence* of that structure or set of items.

[23]For more on this distinction, see e.g. (Fankhauser and Dürr 2021) and references therein. The

through, for example, a gravitational gradiometer (Misner, Thorne, and Wheeler 1973, pp. 400–3). This is a particular device which—when used correctly, i.e., according to some determinate operational prescription—can read out particular components of this tensor. Nevertheless, in virtue of the existence of TEGR, neither spacetime curvature nor spacetime torsion is straightforwardly measurable *per se*.

Given the possibility (indeed, actuality) of the gravitational gradiometer, it is clearly possible to operationalise spacetime curvature in GR. The question, then, is whether it is likewise possible to operationalise torsion in TEGR. If not, then one might argue that the theory is insufficiently interwoven with our actual practice of science. The problem to be addressed, then, is this:

> **Problem 4: Operationalisability:** There is no obvious existing operationalisation of spacetime torsion in TEGR. To the extent that this is true, TEGR is deficient as compared with GR, where operationalisations of spacetime curvature are available.

Is the problem of operationalisability a real problem for TEGR? *Prima facie* yes— although, in fact, in our view the problem can be overcome fairly straightforwardly. To see this, we first consider an intuitive picture of what operationalisation of torsion looks like, quite apart from the context of the whole physical theory, similar to how we teach basic operationalisation of curvature. Then we give two more specific potential operationalisations of spacetime torsion in TEGR. (Of course, ultimately, further operationalisations might also become available.)

An intuitive way to operationalise torsion is as follows. To test whether a given space exhibits the effects of torsion: hold your arms in a ninety degree angle with respect to each other and send two little messengers along the respective directions in which your arms are pointing. Both messengers are equipped with the same notion of length and instructed to walk a certain distance $L$, after the completion of which they each take a right turn (the one in line with your right arm turns left, the one in line with your left arm turns right), after which both messengers again walk the distance $L$.[24] If your messengers end up at the same point, then—*ceteris paribus*—the space does not generically exhibit torsion. If the space does have torsion, the messengers will generically *not* end up in the same spot.[25] That is, if the above procedure is repeated with slight variations of directions, not arriving at the same spot is a necessary (but not sufficient) condition for there being spacetime torsion.

Beyond this intuitive picture, our first further operationalisation of torsion in TEGR 'piggybacks' on the gravitational gradiometer introduced above. Recall (Baha-

---

distinction between empirical significance and operational significance is also discussed by e.g. Chen & Read (2023).

[24] Note that this particular operationalisation is predicated upon the presence of a certain conformal structure and a notion of length (introduced either via the metric or the tetrads, or some other way), i.e., we assume to already know what it means to take a right turn and walk distance $L$.

[25] This is not to say that such operational procedures will *necessarily* adjudicate between torsionful and curved spaces. Indeed, there are possible worlds where the effects of curvature and torsion do not come apart at the operational level, given dynamical effects that precisely offset the manifestations of one over the other—indeed, if the Einstein-Hilbert action is exactly correct, that is precisely the actual world.

monde et al. 2023, Section 2.1.3) that curvature and torsion of a connection are related via the Bianchi identities, as:

$$R^{\mu}{}_{[\nu\rho\sigma]} = \nabla_{[\nu} T^{\mu}{}_{\rho\sigma]} + T^{\mu}{}_{\lambda[\nu} T^{\lambda}{}_{\rho\sigma]}. \tag{4.1}$$

Recall also that there is a straightforward formula relating two Riemann tensors in terms of their difference tensors—see (Malament 2012, problem 1.8.1); here, the salient difference tensor is of course the contorsion tensor (2.8). In that case, one sees that the gravitational gradiometer will also be capable of reading out torsion components and their derivatives. Such read-outs will not be as straightforward as for curvature, for (4.1) shows that components of a curvature tensor correspond to components of a torsion tensor *and their derivatives*; nevertheless, this is sufficient to demonstrate that there still exists a correspondence between the components of torsion and the readouts of actually constructed devices which is at least reasonably straightforward.

Our second potential operationalisation of spacetime torsion in TEGR appeals to considerations presented by Hehl (1971). There—fairly straightforwardly—the author considers the equation of motion governing a test particle with spin, and shows that the components of said equation can be correlated to the components of the torsion tensor—so "we have for the first time a method which allows us in principle to measure all components of the torsion tensor of space-time" (Hehl 1971, p. 226).[26] Of course, it remains to be shown how one would *construct* a device capable of correlating its readouts with torsion components—in this sense, the connection between the relevant geometric object (here the torsion tensor) and device readouts is still less direct than in the above-discussed case of the gradiometer. Nevertheless, we take this to be a *prima facie* good case for the claim that components of the torsion tensor can be put into fairly direct contact with the empirical, and so in this sense are 'operationalisable'.

One might have here a residual question—namely: if torsion and curvature are both operationalisable but not *measurable* (in the above-defined sense), then what becomes of experimental tests of GR, both classic and modern? After all, one test of GR is the famous perihelion shift of Mercury's orbit, which relies on metrical structure; other tests, like gyroscopic frame-dragging, are probed by surveying affine structure.[27] Given empirical equivalence, it is of course to be expected that classic and modern tests of GR deliver the same results in TEGR, albeit that they are differently interpreted, as outlined in detail by Wolf et al. (2024). But it is worth emphasising how special this case is, and that not just *any* manifestation of torsion can be modelled by curvature, nor *vice versa*. One could engage in a thought experiment of imagining a world where a spacetime is torsionful, and not curved, and this fact does indeed manifest at the level of measurable quantities. That is, the torsional terms of the action in this world combine in such a way that they cannot be modelled with a curvature tensor (that there is no dynamically equivalent action that uses only the Riemann tensor and the Levi-Civita connection). Likewise, other worlds can be imagined in which the situation is reversed in favour of the Riemann tensor.[28] However, if our actual world is one in which the dynamical equivalence discussed in Section 2 does indeed hold, then we can express torsion components in terms of curvature components or *vice versa*.[29]

---

[26] Note that Hehl is working within the context of a Riemann-Cartan geometry with torsion.

[27] We thank two anonymous reviewers for emphasising this point.

[28] Some $f(R)$ versus $f(T)$ theories would suffice to illustrate our point here.

## 5 Visualisability

We turn now to our second potential novel problem for torsionful theories (*a fortiori* TEGR), which is whether spacetime torsion can be visualised, and if it cannot be visualised, whether this spells doom for a realist and literalist interpretation of torsionful spacetime theories. More formally, we can state the problem like this:

> **Problem 5: Visualisability**. Curvature as a property of spacetime is easier to visualise than torsion as a property of spacetime. To the extent that visualisability is important for theory choice, GR and TEGR are thus not on a par, thereby breaking any putative underdetermination.

This is a problem for the scientific realist who takes visualisability to be an important selection criterion. Whether one finds this compelling will depend upon one's other commitments in the philosophy of science—for example, logical empiricists will certainly not grant that visualisability bears on scientific knowledge. But for those who think visualisability *is* important for being a realist about a given scientific theory (and there are many such individuals, as we discuss below), the problem of visualisability is certainly relevant to the supposed underdetermination of GR and TEGR.

In this section, we argue that torsion (and hence TEGR *qua* torsionful theory) has no visualisability problems over and above those encountered for curvature in GR. A distinction between 'extrinsic torsion' and 'intrinsic torsion' will prove helpful. Just like extrinsic curvature, extrinsic torsion is visualisable straightforwardly, through a higher-dimensional embedding. There may remain the worry that intrinsic torsion is not visualisable. Yet, even in this case we do not regard the problems for TEGR in this regard as being any more sticky than those for GR, for intrinsic torsion is either (1) just as hard to visualise as intrinsic curvature (Reichenbach argues that it simply takes conscientious training) or (2) just as impossible (following Kantian inconceivability arguments).[30]

### 5.1 Visualisability as epistemically virtuous?

Broadly construed, 'visualisability' consists in the ability to represent something mentally in a visual way, through the mental capacity to form, in the mind's eye, images,

---

[29]Continuing on from the previous footnote: is very possible that even in our actual world the equivalence between the theories is broken, and that at higher energies the effects of torsion and curvature come apart, which could be a world governed by some $f(R)$ theory, or any $f(R, T, Q)$ theory, cf. (Beltrán Jiménez, Heisenberg, and Koivisto 2019; Bahamonde et al. 2023; Chakrabortty, SK, and Kumar Sanyal 2023; Heisenberg 2024). Until we have evidence that this might be so, we assume the actual world is one in which the equivalence holds.

[30]We are careful not to argue that visualisability is necessarily static or changeable. Although in Section 5.3 we consider a position according which visualisability is changeable, the argument we make—that TEGR poses no special threat to visualisability beyond GR—does not hinge on this.

pictures, or representations of objects, phenomena or concepts.[31]

Epistemologically, visualisability raises questions about the reliability and limitations of knowledge obtained by visual means, whether we can or should trust our visual experiences to accurately represent the external world, and, if we cannot, what the limitations of visual perception for accessing the physical world are. Although few deny that visual perception and imagination shape our understanding of the physical world, many doubt whether visualisability is a guide to ontology or a good criterion for theory choice. This is especially so when such visualisation extends from the domain of everyday phenomena into a theoretical or unobservable domain.[32]

Before considering the visualisability of spacetime torsion specifically, we add some motivation to take the criterion seriously. Visualisability is often appealed to as a serious component in the development (discovery) or assessment (justification) of scientific theories. Historically, for example, Minkowski took visualisability as a central tool to get grip on physical reality—as outlined convincingly by Galison (1977). Minkowski claimed that visual-geometric intuition (*geometrische Anschauung*) not only helps us to discover new theorems (his example is number theory), but also gives us insight into physical reality, such as the true geometrical structure of the world. He saw geometry not as mere mathematics or as an abstract formulation of phenomena, but believed that the world was indeed literally a four-dimensional, Lorentzian manifold, and that visualisation played a key role in underpinning that claim. Another salient historical example would be that the early (and persisting) objections to quantum mechanics as describing physical reality—for example by Schrödinger (de Regt 1997) and by Weyl and Einstein (Wolff 2015)—rested on problems with visualisability (*Anschaulichkeit*).

In the philosophy of symmetry, there is a debate between 'interpretationalism' and 'motivationalism', which regards the question as to when two symmetry-related models of a given theory can be regarded as representing the same physical state of affairs (Møller-Nielsen 2017; Martens and Read 2020; Read and Møller-Nielsen 2020; Luc 2023). According to interpretationalism, one may do this *ab initio*; according to motivationalism, one may only do this once one has secured a 'metaphysically perspicuous characterisation' of the common ontology of those symmetry-related models. However, the notion of perspicuity is clearly vague. The English word signals a lack of obscurity, or to have a clear picture of what is going on.[33] To our mind, this notion is tied up intimately with notions from the scientific understanding literature and to a large extent the ability to *visualise* the ontology that underlies the symmetry-related models, making them intelligible. The thought, then, would be that if models of TEGR are not visualisable, then they are not perspicuous, and so in turn are not amenable to a scientific realist treatment. To the extent that visualisability will play a role in this debate, we believe the following discussions to be thoroughly motivated.

---

[31]What does 'in a visual way' mean? We suggest that it means representing something using visual elements (such as images, diagrams, charts). It means presenting information or concepts in a manner that can be seen and understood through the sense of sight, rather than relying on, for example, written or verbal descriptions. It is that which people with aphantasia cannot do.

[32]Here there is a spectrum, with those more positivistically-inclined towards the one end and (arguably) with 'primitive ontologists' (see e.g. Allori 2015) towards the other.

[33]Read (2024) cashes out the notion in terms of psychological satisfaction.

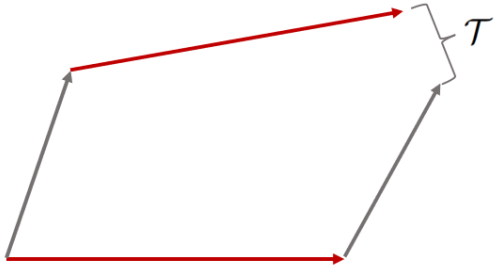 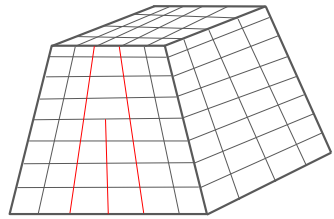

Figure 1: The torsion tensor is a measure of the non-closure of parallelograms.

Figure 2: Crystal structure with edge dislocation.

## 5.2 Extrinsic torsion

Visualisations of geometry are likely to be the most common visualisation in all of science, starting with the cubes and spheres in primary school geometry. Such pictures are embedded within our three-dimensional Euclidean surroundings. Some years later, such embedded shapes become the way in which one is first introduced to curvature in GR classes. The most universally taught visualisation of curved manifolds in textbooks of GR, indeed, is a sphere over which vectors are parallel transported. Following segments of great circles from the North pole to the equator, then over the equator, and then back to the North pole, indicates the presence of curvature on this two-dimensional manifold—consider the famous cover of Misner, Thorne and Wheeler's *Gravitation* (1973). Only later does one move to the more abstract characterisation of the *intrinsic* curvature of a differentiable manifold. The sphere's curvature is assessed extrinsically because it is visualised as being embedded in a three-dimensional Euclidean space.

Given that the visualisation of extrinsic *curvature* is relatively straightforward, one might worry that torsion is somehow metaphysically problematic without an analogous visualisation. Such visualisations of torsion, however, already exist, although they are less well-known. Here, we focus on two such visualizations, following Figures 1 and 2: respectively, (i) the non-closure of parallelograms, and (ii) the application of torsional geometry to crystal structures.

First, the usual picture that one is presented with when torsion is discussed is that of two vectors at an angle from each other, which are then parallel transported along each other (Figure 1). In a torsion-free space, such an operation would result in 'the closure of the parallelogram', or, more simply, the shape visualised would be a parallelogram. When the space has torsion, however, after a vector **u** is transported over a vector **v** and **v** is transported over **u**, there will be a gap: the torsion tensor is a measure of this gap.[34]

Even though this picture should be sufficient to get to the core of what torsion amounts to, at first sight one might object to it as being unnatural or hard to interpret. At what, for example, are these transported vectors pointing? The uneasiness signalled by this question occurs only when one forgets about the background points:

---

[34] As mentioned earlier, strictly speaking we are dealing here with infinitesimal parallelograms.

the transported vectors are simply indicating other points than the ones they would have both indicated were there no torsion.

Second, the application of torsion to crystalline structures has been studied extensively—see e.g. (Lazar 2001; Hehl and Obukhov 2007; Lazar and Hehl 2010). Because torsion introduces a mismatch in orientation from lattice point to lattice point, it is particularly well-suited for studying crystal defects and dislocations. One example is shown in Figure 2, and we strongly recommend (Hehl and Obukhov 2007) for many more examples and exhaustive analysis of how such structures come about. The crystal structures thus produced make for good indirect visualisations of torsion itself, if one regards the lattice sites not as *on* a manifold, but as points *of* the manifold.[35]

Visualisation of extrinsic torsion thus comes with no special problems that do not come in the case of extrinsic curvature. Not only is the non-closing of parallelograms already a visualisation, three-dimensional visualisations of torsion are routinely applied in condensed matter physics, which seem to readily extent to possible visualisations of the space manifold. One might still worry about the extent to which this can be generalised to intrinsic torsion. In the next subsection, we argue that the visualisation of curvature suffers from the same obstacles and follow Reichenbach into thinking that the situation for visualising unintuitive geometries may not be as bad as one might initially think.

## 5.3 Intrinsic torsion

Already in the case of curvature, opinions differ over whether 'intrinsic' (i.e., 'immanent'; 'from within') visualisations of non-Euclidean geometries are possible. After all, any explicit drawing of anything on a piece of two-dimensional paper (or an explicit arrangement of visual cues in a three-dimensional room) will, for all practical purposes, be embedded in a flat space. To visualise without a piece of paper—in the mind's eye, so to speak—does not appear to alleviate the need for an embedding space. On the one hand, we have the Kantian observation that in order to conceive of something at all, situating it in Euclidean space is a prerequisite. Indeed, visualisations in the previous subsection of curvature as a two-sphere or torsion as a three-dimensional crystal structure are embedded in a flat three-space. To some of a traditional Kantian inclination, visualising intrinsic curvature without a Euclidean embedding space might seem impossible.[36]

Hasok Chang, in the context of intelligibility, suggests that the whole activity of visualisation is intelligible only insofar as it is set in Euclidean space:

> Visualization, I think, is supported by a set of principles that form the

---

[35]In addition to these examples, there are yet further ways to visualise torsion—for example, in terms of the twisting of screws, on which (in two dimensions) see, e.g., https://www.youtube.com/watch?v=1YTKedLQOa0.

[36]What we have in mind is those who, in Kantian tradition, hold that our perceptual judgements necessitate a Euclidean space; after all, it was one of Kant's main examples of the transcendental deduction that space must be Euclidean *because* we cannot imagine otherwise. We will not know whether Immanuel Kant would have broadened his definition of spatio-temporality to non-Euclidean versions, but here we leave space for those who hold on to the narrower, Euclidean definition.

basis of Euclidean geometry. (This explains why non-Euclidean geometry is deeply unintelligible to those of us who try to visualize what is going on there.) (2009, p. 78)

To the extent that this is correct, the problem of visualisation gets no traction on TEGR, as neither intrinsic torsion nor intrinsic curvature will be visualisable in this case.

On the other hand, although visualisation of intrinsic curvature of spacetime is unintuitive—in the sense that it is far removed from everyday practice—it may nevertheless be *trained*. One gets such an impression from (Norton 2007, Chapter 24), or from Jeffrey Weeks' book (2001), and of course Reichenbach (1928) argues for this explicitly. Right after dismissing—in light of the empirical success of relativity theory—the Kantian view that Euclidean geometry is epistemically *a priori*, he (pp. 31–2) moves on to discuss what he calls the 'visual *a priori*'. This is a methodological rule that prefers Euclidean geometry over other geometries on the basis of one's ability to visualise as an "innate property of the human mind". Reichenbach (§9–§11, pp. 37–58 and §13, pp. 81–92) then proceeds on the topic of visualisation for four (!) sections. In fact, the entire context of his well-known relativity of geometry, as stated by the (infamous) *Theorem θ*, is the ultimate failure of simultaneously adhering to two desirable methodological principles. On the one hand there's the visual *a priori*, while on the other there is a 'principle of elimination of universal forces'. Theorem $\theta$ posits that *any* geometry $G$ can be swapped for another geometry $G'$ if compensated for by a so-called universal force $F$. Such a force is universal in that it cannot be screened off by insulating walls and acts on all matter in the same way. Thus a universal force affects all measuring instruments—rods and clocks—in the same way; as such, only the combination $G + F$ is empirically accessible. This is the reason why the visual *a priori* is read as a *methodological* rule: according to Reichenbach, it is a matter of convention which geometry we use.

As such, the universal force can be accommodated by a suitable conventional choice. The 'principle of elimination of universal forces' is one such choice. Thus, one would be left with $G + 0$ and the geometry can be directly surveyed by rods and clocks. This, however, clashes with the alternative choice that satisfies the rule of the visual *a priori*, which is: "The comparison of length is to be performed in such a way that Euclidean geometry will be the result" (Reichenbach 1928, p. 34). In this case, one is left with $G_{\text{Euclid}} + F$, which should be easy to visualise, at the cost of introducing these mysterious universal forces. Reichenbach is careful to state that this does not inform us about the "space of real objects", as it only accommodates the *epistemological function of visualisation*: to ground subjective preference.

In the end, Reichenbach decides against the visual *a priori*, at the cost of easy visualisability. The reason is not just the above observation that subjective preferences do not inform the reality of space; it is also that Reichenbach, all things considered, *does not believe that non-Euclidean geometries are as unvisualisable as they are commonly taken to be.* For example, in Euclidean geometry a figure can be scaled up from smaller to larger without changing shape (imagine a glass marble grow to the size of a basketball to the size of one of Brussels' Atomium's iron atoms). In non-Euclidean geometry the relations *do* change: the sum of the angles of a triangle and the ratio between the circumference and the diameter of a circle depend on the absolute size of the figure. Even though this hinders easy visualisation, Reichenbach (1928, p. 45) proposes that one should simply modify the method of visualisation: in addition to visualising the shape, the "smaller figures must be imagined as distorted."

Turning to intrinsic curvature, Reichenbach argues that we have insufficiently *experienced* non-Euclidean geometries and that our ability to visualise is adjusted accordingly. This ability, however, is not set in stone: with sufficient practice we can learn to visualise the non-Euclidean just as well. The untrained automobile driver (Reichenbach 1928, p. 55) who uses a convex mirror to keep track of the road behind, will see the images as distorted, changing shape; with sufficient driving experience, however, these changes in shape are no longer experienced as distortions, and the picture is taken as it presents itself in a non-Euclidean way.

The reason why we can adjust our ability to visualise, according to Reichenbach, is because of our command of *logical capacity*, which we can use to adopt different definitions of congruence. The principle that we are accustomed to in our apparently Euclidean world is, in mathematical terms, that "[...] concepts of *line of equal distance from a given straight line* and of *straightest line* are coextensive" (Reichenbach 1928, p. 57, our emphasis). Such a definition of congruence is not logically necessitated and, indeed, letting go of this particular definition allows us to "emancipate" ourselves from our "native" Euclidean geometrical intuitions, rather analogous to learning a language other than one's native language. Thus, in visualising curvature we need to internalise the following lesson:

> **Non-Euclidean lesson**. *Affine geodesics* are not necessarily congruent with the lines of equal distance from a given straight line.

Although Reichenbach's main example concerns the visualisation of parallel lines in Bolyai-Lobachevsky geometry, his presentation remains more of a proof of concept that a visualisation need not always be Euclidean than a constructive alternative definition of congruence. The alternative definition of congruence which he proposes is simply the denial of the Euclidean definition of congruence: "We have visualized the interior [i.e., intrinsic] curvature, since interior curvature is nothing but the deviation from Euclidean congruence" (Reichenbach 1928, p. 57).

To the extent that someone holds that intrinsic curvature can be visualised in such ways, we likewise think that intrinsic torsion can be visualised. Those familiar with GR have internalised the Non-Euclidean lesson very well. Within Non-Euclidean spaces however, there are of course many concepts to fine-grain further. To achieve the visualisation of intrinsic torsion, then, we need to use our logical capacity to rid ourselves of the common convictions that *straightest curves* are coextensive with *shortest curves*. Or in relativistic language:

> **Teleparallel lesson**. *Affine geodesics* (the straightest lines) are not necessarily congruent with the *extremal curves* (the lines with extremal distance), where the former is given by the affine connection and the latter by the metric.

All we have to do is get used to it: through practice one can become proficient in the new language and come to visualise intrinsic torsion.

Thus we see no obstacles for visualising torsion, either intrinsically or extrinsically—at least not in addition to any obstacles that stand in the way of visualising curvature. The only way we see to keep the problem of visualisation alive is to point out that pictures of torsion are relatively underdeveloped. We are, however, optimistic that sufficient practice and acquaintance can bring torsion quite literally into focus.

# 6 Conclusion

The theme throughout this paper has been that the arguments against TEGR which we have encountered are typically based on a familiarity with GR which will lose their appeal as one gains more familiarity with teleparallel gravity. TEGR is not easily dismissed as a viable alternative to GR on the basis of (existing) arguments that appeal to its intertranslatibility with GR (Section 3.1), its surplus (gauge) structure relative to GR (Section 3.2), any inability to provide acceptable inertial structure (Section 3.3), any lack of operationalisability (Section 4), or on any problems with visualisability that exceed those of GR (Section 5).

As scientific realists, we recognise that our conclusion is negative: we would like to make the strongest possible justified claims about what nature is like, but a situation of underdetermination of GR and TEGR weakens such pretensions. It takes away from us the justification to assert confidently that GR tells us that spacetime is curved. After all, spacetime could also be torsionful if TEGR is correct. When responding to cases of underdetermination of this kind there is always available a variety of responses; to canvas here just some:

1. One can hold that even though one of the theories is correct, we cannot know which one, resulting in a (transitory or in-principle) anti-realism or agnosticism.

2. One can adopt conventionalism, according to which geometrical propositions do not *per se* have truth values, but can be assented to as a matter of free choice, with each alternative epistemically as good as any other. (See (Dürr and Read 2024).)

3. One can argue that empirical justification is not our only source of underpinning the truth, and we need to appeal to something *supra*-empirical. General examples of such criteria are coherence, simplicity, fruitfulness, explanatory power; various physics-specific examples are locality, determinism, or invariance under a certain symmetry group.[37]

As scientific realists, we also recognise that we cannot let our desire to remove underdetermination lead us astray, into a territory where our ontological commitments are based on a familiarity with textbook theories, or by a philosophy so strong that we can no longer see the ontology clearly. As a fourth route, adding to the above, one may attempt to reconcile the ontological claims of both theories by abstracting away from torsion and curvature specifically, and looking at what they share in common—a project left for future research (Mulder 2024a; Wolf, Sanchioni, and Read 2024). Alternatively, if one nevertheless finds good other reasons to dismiss teleparallel gravity as a viable alternative this may well help to formulate a philosophical framework to help us recognise ontology for strongly underdetermined theories in general—a framework hopefully more potent than our current ones.

---

[37]We thank an anonymous reviewer on pressing us to give an overview of the available options in cases of underdetermination.

## Acknowledgements

We are very grateful to Jeremy Butterfield, Hasok Chang, Erik Curiel, Neil Dewar, Ronnie Hermens, Eleanor Knox, Peter Morgan, and William Wolf for illuminating discussion and assessment of earlier drafts, which improved the paper significantly. We also thank four reviewers for deep engagement with the manuscript and many comments that have improved the arguments and made them easier to follow. R.M. is grateful to the Tarner Scholarship in Philosophy of Science and History of Scientific Ideas, Trinity College, Cambridge; J.R. is grateful to the Leverhulme Trust for support.